\documentclass[12pt]{article}

\usepackage{amsmath}
\usepackage[dvips]{epsfig}

\allowdisplaybreaks[4]

\renewcommand{\title}[1]{\null\vspace{10mm}\noindent
                         {\Large{\bf #1}}\vspace{10mm}}
\newcommand{\authors}[1]{\noindent{\large #1}\vspace{20mm}}
\newcommand{\address}[1]{{\center{\noindent\small\itshape #1\vspace{0mm}}}}

\makeatletter
\def\section{\@startsection{section}{1}{\z@}{-3.25ex plus -1ex minus
             -.2ex}{1.5ex plus .2ex}{\normalfont\bfseries}}
\def\subsection{\@startsection{subsection}{1}{\z@}{-3.25ex plus -1ex
                minus -.2ex}{1.5ex plus .2ex}{\normalfont\itshape}}

\renewenvironment{thebibliography}[1]
         {\section*{References}\frenchspacing\small
          \begin{list}{[\arabic{enumi}]}
         {\usecounter{enumi}\parsep=2pt\topsep 0pt
         \settowidth{\labelwidth}{[#1]}
         \leftmargin=\labelwidth\advance\leftmargin\labelsep
         \rightmargin=0pt\itemsep=0pt\sloppy}}{\end{list}}

\makeatother

\def\smallhatbarc{\raisebox{-1mm}{\scriptsize$\hat{\bar{%
                  \raisebox{1mm}{\scriptsize$c$}}}$}}

\newtheorem{thm}{Theorem}

\begin{document} 
 
\thispagestyle{empty}

\begin{titlepage}

\begin{center}
\hspace*{\fill}{{\normalsize \begin{tabular}{l}
                              {\sf hep-th/0108045}\\
                              {\sf REF. TUW 01-019}\\
                              {\sf REF. UWThPh-2001-32}
                              \end{tabular}   }}

\title{Noncommutative Lorentz Symmetry \\[3mm]
and the Origin of the Seiberg-Witten Map}

\authors {  \large{A.~A.~Bichl$^{1}$, J.~M.~Grimstrup$^{2}$, 
H.~Grosse$^{3}$, E.~Kraus$^{4}$,\\[1mm] 
L.~Popp$^{5}$, M.~Schweda$^{6}$, R.~Wulkenhaar$^{7}$ }}  

\vspace{-15mm}
       
\address{$^{1,2,5,6}$  Institut f\"ur Theoretische Physik, 
Technische Universit\"at Wien \\
Wiedner Hauptstra\ss e 8--10, A-1040 Wien, Austria}
\address{$^{3,7}$  Institut f\"ur Theoretische Physik, 
Universit\"at Wien\\
Boltzmanngasse 5, A-1090 Wien, Austria}
\address{$^{4}$ Physikalisches Institut der Universit\"at Bonn,\\
          Nu\ss{}allee 12, D-53115 Bonn, Germany }

\footnotetext[1]{bichl@hep.itp.tuwien.ac.at, work supported in 
part by ``Fonds zur F\"orderung der Wissenschaftlichen Forschung'' 
(FWF) under contract P14639-TPH.}

\footnotetext[2]{jesper@hep.itp.tuwien.ac.at, work supported by 
The Danish Research Agency.}

\footnotetext[3]{grosse@doppler.thp.univie.ac.at.}

\footnotetext[4]{kraus@th.physik.uni-bonn.de.}

\footnotetext[5]{popp@hep.itp.tuwien.ac.at, work supported in part 
by ``Fonds zur F\"orderung der Wissenschaftlichen Forschung'' (FWF) 
under contract P13125-PHY.}

\footnotetext[6]{mschweda@tph.tuwien.ac.at.}

\footnotetext[7]{raimar@doppler.thp.univie.ac.at, Marie-Curie Fellow.}

\vspace{10mm}

\begin{minipage}{12cm}
  
  {\it Abstract.} We show that the noncommutative Yang-Mills field
  forms an irreducible representation of the (undeformed) Lie algebra
  of rigid translations, rotations and dilatations. The noncommutative
  Yang-Mills action is invariant under combined conformal
  transformations of the Yang-Mills field and of the noncommutativity
  parameter $\theta$. The Seiberg-Witten differential equation results
  from a covariant splitting of the combined conformal transformations
  and can be computed as the missing piece to complete a covariant
  conformal transformation to an invariance of the action.
  \vspace*{1cm}
\end{minipage}

\end{center}

\end{titlepage}

\section{Introduction}

In noncommutative field theory one of the greatest surprises is the
existence of the so-called Seiberg-Witten map \cite{Seiberg:1999vs}.
The Seiberg-Witten map was originally deduced from the observation
that different regularization schemes (point-splitting vs.\ 
Pauli-Villars) in the field theory limit of string theory lead either
to a commutative or a noncommutative field theory and thus suggest an
equivalence between them.

A particular application of the Seiberg-Witten map is the construction
of the noncommutative analogue of gauge theories with arbitrary gauge
group, which automatically leads to enveloping algebra-valued fields
involving infinitely many degrees of freedom \cite{Jurco:2000ja}. The
Seiberg-Witten map solves this problem in an almost miraculous manner
by mapping the enveloping algebra-valued noncommutative gauge field to
a commutative gauge field with finitely many degrees of freedom. 

The renormalization of noncommutative Yang-Mills (NCYM) theories is an
open puzzle: Loop calculations \cite{Matusis:2000jf} and
power-counting analysis \cite{Chepelev:2001hm} show the existence of a
new type of infrared divergences. The circumvention of the infrared
problem by application of the Seiberg-Witten map leads to a
power-counting non-renormalizable theory with infinitely many
vertices. In an earlier work \cite{Bichl:2001cq} we have proven the
two-point function of $\theta$-expanded noncommutative Maxwell theory
to be renormalizable to all orders. However, to show renormalizability
of all $N$-point functions one cannot proceed without strong
symmetries that limit the number of possible counterterms. In
particular, one needs to find a symmetry that fixes the special
$\theta$-structure of the $\theta$-expanded theory.

The intuition that the symmetry searched for is related to space-time
symmetries leads us to an investigation of rigid conformal symmetries
(translation, rotation, dilatation) for NCYM theory characterized by a
constant field $\theta^{\mu\nu}$. The term rigid means that the factor
$\Omega$ in the conformal transformation $(ds')^2=\Omega^2 ds^2$ of
the line element is constant. The reason for this restriction is that
$\theta$ has to be constant in all reference frames.

We show in this paper that the noncommutative Yang-Mills field
$\hat{A}$ forms an irreducible spin-$1$ representation of the
\emph{undeformed} Lie algebra of conformal transformations.
We also prove that the noncommutative Yang-Mills (NCYM) action
is invariant under the sum of the conformal transformations of
$\hat{A}$ and of $\theta$. This result can either be regarded as an
exact invariance (compatible with gauge
transformations) with respect to \emph{observer Lorentz transformations}
or as the quantitative amount of symmetry breaking under
\emph{particle Lorentz transformations}, see also Section~\ref{sec3}.

Regarding the combined conformal transformations of $\hat{A}$ and
$\theta$, one can consider various splittings into individual
transformations. There is one (up to gauge transformations)
distinguished splitting for which both individual components are
compatible (covariant) with gauge transformations, i.e.\ the
commutator of these components with a gauge transformation is again a
gauge transformation. Whereas the $\theta$-part of this covariant
splitting cannot be computed, the $\hat{A}$-part is easily constructed
by a covariance ansatz involving \emph{covariant coordinates}
\cite{Cornalba:2000ah,Madore:2000en}. This covariance ansatz
generalizes the gauge-covariant conformal transformations which in its
commutative form were first investigated by Jackiw
\cite{Jackiw:1978ar,Jackiw:1980}. These transformations are loosely
related to the improvements allowing to pass from the canonical
energy-momentum tensor to the symmetric and traceless one. Now, the
covariant $\theta$-complement of the covariant transformation of
$\hat{A}$ can easily be computed as the missing piece to achieve
invariance of the NCYM action. The result is the Seiberg-Witten
differential equation \cite{Seiberg:1999vs}.

Almost all splittings of the combined conformal transformation of
$\hat{A}$ and $\theta$ lead to a first-order differential equation for
$\hat{A}$ which can be used to express the noncommutative fields in
terms of initial values living on commutative space-time. The
covariant splitting (which leads to the Seiberg-Witten differential
equation) has the distinguished property that the resulting
$\theta$-expansion of a gauge-invariant noncommutative action is
invariant under commutative gauge transformations. This was the
original motivation for the Seiberg-Witten map. We would like to point
out, however, that the original gauge-equivalence condition
\cite{Seiberg:1999vs} is more restrictive than the approach of this
paper---a fact made transparent by our investigation of noncommutative
conformal symmetries. Moreover, we prove that the $\theta$-expansion
of the noncommutative conformal symmetries reduces to the commutative
conformal symmetries.

All this means that there are two quantum field theories associated
with the NCYM action. The first one is obtained by a direct
gauge-fixing of the NCYM action and the other one by gauge-fixing of
the $\theta$-expanded NCYM action. The second approach was adopted in
\cite{Bichl:2001nf,Bichl:2001cq}: Take the Seiberg-Witten expansion of
the NCYM action as a very special type of an action for a commutative
gauge field $A_\mu$ coupled to a constant external field
$\theta^{\mu\nu}$ and quantize it in the ordinary way (with the linear
gauge-fixing in \cite{Bichl:2001nf}). It is not completely clear in
which sense this is equivalent to the first approach of a direct
quantization of the noncommutative Yang-Mills action. The infrared
problem found in noncommutative quantum field theory
\cite{Matusis:2000jf,Chepelev:2001hm} and its absence in the approach
of \cite{Bichl:2001nf} shows the inequivalence at least on a
perturbative level. For interesting physical consequences of the
Seiberg-Witten expanded action in noncommutative QED see
\cite{Guralnik:2001ax}.

The paper is organized as follows: First we recall in Section
\ref{sec2} necessary information about noncommutative field theory and
covariant coordinates. In Section \ref{sec3} we distinguish between
observer and particle Lorentz transformations. After a review of rigid
conformal symmetries in the commutative setting in Section \ref{sec4}
we extend these structures in Section \ref{sec5} to noncommutative
Yang-Mills theory, deriving in particular the Seiberg-Witten
differential equation and the $\theta$-expansion of the noncommutative
conformal and gauge symmetries. In Section~\ref{sec7} we comment on
quantization and Section~\ref{sec8} contains the summary. Longer but
important calculations are delegated to the Appendix.

\section{Noncommutative geometry and covariant coordinates} 
\label{sec2}

In this section we give a short introduction to noncommutative field
theory and the concept of covariant coordinates.  We consider a
noncommutative geometry characterized by the algebra 
\begin{equation}
\left[\mathsf{x}^\mu ,\mathsf{x}^\nu \right] = \mathrm{i} \theta^{\mu\nu},
\label{x}
\end{equation}
where $\theta^{\mu\nu}$ is an antisymmetric constant tensor. The
noncommutative algebra may be represented on a commutative manifold
by the $\star$-product
\begin{equation}
(f\star g)(x)=\int \frac{d^4 k}{(2\pi)^4}
\int \frac{d^4 p}{(2\pi)^4} \,\mathrm{e}^{-\mathrm{i}
(k_\mu +p_\mu )x^\mu } \,
\mathrm{e}^{-\frac{\mathrm{i}}{2}\theta^{\mu\nu} k_\mu p_\nu }
\tilde{f}(k)\,\tilde{g}(p)~,
\label{ny-1}
\end{equation}
where $f(x)$ and $g(x)$ are ordinary functions on Minkowski space and
$\tilde{f}(p)$ and $\tilde{g}(p)$ their Fourier transforms. Denoting the
ordinary (commutative) coordinates by $x$ we have
\begin{equation}
[x^\mu ,x^\nu ]_\star \equiv x^\mu \star x^\nu -x^\nu \star x^\mu 
= \mathrm{i} \theta^{\mu\nu} .
\label{alg}
\end{equation}
Let us now consider an infinitesimal gauge transformation $\delta^G$ of
a field $\Phi(x)$,
\begin{equation}
\delta^G \Phi(x) = \mathrm{i}\epsilon(x)\star\Phi(x)~,
\end{equation}
with $\epsilon(x)$ being an infinitesimal gauge parameter. As usual 
one chooses the coordinates to be invariant under gauge
transformations, $\delta^G x = 0$.  However, with this construction
one finds that multiplication by $x$ does not lead to a covariant
object:
\begin{equation}
\delta^G \big(x^\mu \star \Phi(x)\big) \neq \mathrm{i} \epsilon(x) \star 
\big(x^\mu \star \Phi(x) \big)~.
\label{not}
\end{equation}
The solution of this problem, which was given in \cite{Madore:2000en},
is to introduce {\it covariant coordinates} \cite{Cornalba:2000ah}
\begin{equation}
\hat{X}^\mu \equiv x^\mu \mathbf{1} +\theta^{\mu\nu}\hat{A}_\nu~,
\label{covc}
\end{equation} 
where the transformation of the field $\hat{A}(x)$ is defined by the
requirement
\begin{equation}
\delta^G \big(\hat{X}^\mu \star \Phi(x)\big) 
= \mathrm{i} \epsilon(x) \star \big( \hat{X}^\mu \star \Phi(x)
\big)~. 
\label{ups}
\end{equation}
The relation (\ref{ups}) leads to the transformation rule for the
field $\hat{A}(x)$
\begin{equation}
\delta^G \hat{A}_\mu (x)= \partial_\mu \epsilon (x) 
-\mathrm{i} [\hat{A}_\mu(x),\epsilon(x)]_\star 
\equiv \hat{D}_\mu \epsilon(x)~,
\label{Ahat}
\end{equation}
and $\hat{A}(x)$ is interpreted as a noncommutative gauge field. In
this way gauge theory is seen to be intimately related to the
noncommutative structure (\ref{alg}) of space and time. The covariant
coordinates fulfill
\begin{equation}
\big[\hat{X}^\mu ,\hat{X}^\nu \big]_\star 
= \mathrm{i} \theta^{\mu\nu} + \mathrm{i}\theta^{\mu\alpha}
\theta^{\nu\beta} \hat{F}_{\alpha\beta}~, 
\label{covX}
\end{equation} 
where $\hat{F}_{\alpha\beta} = \partial_\alpha \hat{A}_{\beta} 
-\partial_{\beta} \hat{A}_\alpha  
- \mathrm{i} \big[ \hat{A}_\alpha ,\hat{A}_{\beta}\big]_\star$ is 
the noncommutative field strength.

\section{Observer versus particle Lorentz transformations}
\label{sec3}

In general one should distinguish between two kinds of Lorentz (or
more general, conformal) transformations (see \cite{Colladay:1998fq}
and references therein). Lorentz transformations in special relativity
relate physical observations made in two inertial reference frames
characterized by different velocities and orientations. These
transformations can be implemented as coordinate changes, known as
{\it observer Lorentz transformations}. Alternatively one considers
transformations which relate physical properties of two particles with
different helicities or momenta within one specific inertial frame.
These are known as {\it particle Lorentz transformations}.  Usually
(without background) these two approaches are equivalent. However, in
the presence of a background tensor field this equivalence fails,
because the background field will transform as a tensor under
observer Lorentz transformation and as a set of scalars under particle
Lorentz transformations.

Thirdly, having a background tensor field one may
consider the transformations of {\it all} fields within a specific
inertial frame simultaneously, including the background field. These
transformations are known as {\it (inverse) active Lorentz
  transformations} and are equivalent to observer Lorentz
transformations.

What kind of `field' is $\theta^{\alpha\beta}$? Since we are
considering the case of a constant $\theta$, it certainly is a
background field. Therefore, all results of this paper refer to
`observer' transformations. This also matches the setting of
noncommutative field theory appearing in string theory. Here $\theta$
is related to the inverse of a `magnetic field' (mostly taken to be
constant). In this sense, Lorentz
invariance of the action means that its value is the same for
observers in different inertial reference frames. Since invariance of
the action always involves the sum of conformal transformations of
$\hat{A}$ and $\theta$, see Section~\ref{sec51}, one can however take
the `particle' point of view and regard our `observer' invariance
as the quantitative amount of `particle' symmetry breaking due to
the presence of $\theta$.

However, we find it desirable to extend the general analysis to the
case of a non-constant $\theta$. In this case one could choose to view
$\theta$ as a dynamical field which also transforms under `particle'
transformations.

In the rest of the paper we will simply refer to conformal
transformations, leaving out the `observer' prefix.

\section{Rigid conformal symmetries: commutative case}  
\label{sec4}

The Lie algebra of the rigid conformal transformations is generated by 
$\{P_\tau,M_{\alpha\beta},D\}$ and the following
commutation relations:
\begin{align}
[P_\tau,P_\sigma] &=0~, &  [D,D] &=0~, 
\nonumber
\\*
[P_\tau, M_{\alpha\beta}] &= g_{\tau\beta} P_\alpha 
- g_{\tau\alpha} P_\beta ~,
& [P_\tau ,D] &= - P_\tau~,
\nonumber
\\*
[M_{\alpha\beta}, M_{\gamma\delta}] &= 
g_{\alpha\gamma} M_{\beta\delta} 
- g_{\beta\gamma} M_{\alpha\delta} 
- g_{\alpha\delta} M_{\beta\gamma} 
+ g_{\beta\delta} M_{\alpha\gamma} ~,
& [M_{\alpha\beta},D] &=0~.
  \label{cr}
\end{align}
A particular representation is given by infinitesimal rigid conformal 
transformations of the coordinates $x^\mu$, 
\begin{align}
(x^\mu)^T &= (1 + a^\tau \rho_x(P_\tau)) x^\mu  + \mathcal{O}(a^2) ~, & 
\rho_x(P_\tau) &= \partial_\tau  &&  \text{(translation),}
  \label{Tx}
\\
(x^\mu)^R &= (1 + \omega^{\alpha\beta} \rho_x(M_{\alpha\beta})) x^\mu  
+ \mathcal{O}(\omega^2) ~, & 
\rho_x(M_{\alpha\beta}) &= x_\beta \partial_\alpha - x_\alpha \partial_\beta 
&&  \text{(rotation),}
  \label{Rx}
\\
(x^\mu)^D &= (1 + \epsilon \rho_x(D)) x^\mu  
+ \mathcal{O}(\epsilon^2) ~, & 
\rho_x(D) &= - x^\delta \partial_\delta &&  \text{(dilatation),}
  \label{Dx}
\end{align}
for constant parameters $a^\tau,\omega^{\alpha\beta},\epsilon$. 

A field is by definition an irreducible representation of the Lie
algebra (\ref{cr}). In view of the noncommutative generalization we
are interested in the Yang-Mills field $A_\mu$ and the constant
antisymmetric two-tensor field $\theta^{\mu\nu}$ whose representations
are given by
\begin{align}
\rho_1(P_\tau) A_\mu  &= W^T_{A;\tau} A_\mu ~, &
W^T_{A;\tau}  &:= \int d^4x\,\mathrm{tr} 
\Big(\partial_\tau A_\mu \,\frac{\delta}{\delta A_\mu}\Big) ,
  \label{TA}
\\*
\rho_1(M_{\alpha\beta}) A_\mu &= W^R_{A;\alpha\beta} A_\mu ~, &
W^R_{A;\alpha\beta} &:=
\int d^4x\,\mathrm{tr} \Big(
\big(g_{\mu\alpha} A_\beta -g_{\mu\beta} A_\alpha 
+ x_\alpha \partial_\beta A_\mu - x_\beta \partial_\alpha A_\mu \big)
\,\frac{\delta}{\delta A_\mu}\Big),
\label{RA}
\\*
\rho_1(D) A_\mu &= W^D_{A} A_\mu ~, & 
W^D_A &:= \int d^4x\,\mathrm{tr} \Big(
\big( A_\mu + x^\delta \partial_\delta A_\mu\big)\,
\frac{\delta}{\delta A_\mu}\Big),
\label{DA}
\end{align}
and\footnote{The translation invariance
  $\rho_{-2}(P_\tau)\theta^{\mu\nu}=0$ qualifies $\theta^{\mu\nu}$ as a
  constant field. It takes however different (constant!) values in
  different reference frames. The necessity to have a constant field
  in the model forces us to restrict ourselves to rigid conformal 
  transformations. Local conformal transformations as in
  \cite{Kraus:1992cq} are incompatible with constant fields. In particular, 
  the special conformal transformations $K_\sigma$ are excluded because
  the commutator $[K_\sigma,P_\tau] = 2 (g_{\sigma\tau} D -
  M_{\sigma\tau})$ cannot be represented.}
\begin{align}
\rho_{-2}(P_\tau) \theta^{\mu\nu} &= W^T_{\theta;\tau} \theta^{\mu\nu}
& 
W^T_{\theta;\tau}\theta^{\mu\nu} &:= 0 ~,
  \label{TT}
\\
\rho_{-2} (M_{\alpha\beta}) \theta^{\mu\nu} &= 
W^R_{\theta;\alpha\beta} \theta^{\mu\nu} ~, & 
W^R_{\theta;\alpha\beta}\theta^{\mu\nu} &:= 
\delta^\mu_\alpha \theta_\beta^{~\nu} 
- \delta^\mu_\beta \theta_\alpha^{~\nu} 
+ \delta^\nu_\alpha \theta^\mu_{~\beta} 
- \delta^\nu_\beta \theta^\mu_{~\alpha} 
~,
  \label{RT}
\\
\rho_{-2}(D) \theta^{\mu\nu} &= W^D_\theta \theta^{\mu\nu} ~, &
W^D_\theta\theta^{\mu\nu} &:=  - 2 \theta^{\mu\nu} 
~.
  \label{DT}
\end{align}
Throughout this paper we use the following differentiation rule for an
antisymmetric two-tensor field:
\begin{align}
  \label{dt}
\frac{\partial \theta^{\mu\nu}}{\partial \theta^{\rho\sigma}} 
:= \frac{1}{2} \Big(
\delta^\mu_\rho \delta^\nu_\sigma
-\delta^\mu_\sigma \delta^\nu_\rho\Big)~.  
\end{align}
The factor $\frac{1}{2}$ in (\ref{dt}) ensures the same rotational
behaviour of the spin indices in (\ref{RA}) and (\ref{RT}). The
Yang-Mills action
\begin{align}
  \label{YM}
\Sigma = -\frac{1}{4 g^2} \int d^4x\,\mathrm{tr} \big( F_{\mu\nu}
F^{\mu\nu} \big)~,
\end{align}
for $F_{\mu\nu}=\partial_\mu A_\nu - \partial_\nu A_\mu - \mathrm{i}
[A_\mu,A_\nu]$ being the Yang-Mills field strength and $g$ a coupling
constant, is invariant under (\ref{TA})--(\ref{DA}).
Moreover the action (\ref{YM}) is invariant under 
gauge transformations
\begin{align}
  \label{gauge}
W^G_{A;\lambda} = \int d^4x\,\mathrm{tr}\Big( D_\mu \lambda \,
\frac{\delta}{\delta A_\mu} \Big)  ~, \qquad \qquad
D_\mu \bullet = \partial_\mu \bullet - \mathrm{i} [A_\mu ,\bullet]~,
\end{align}
with a possibly field-dependent transformation parameter $\lambda$.

\section{Rigid conformal symmetries: noncommutative case}  
\label{sec5}

In this section we show that the noncommutative gauge field forms an
irreducible representation of \emph{the same undeformed} Lie algebra
of rigid conformal transformations. To obtain the representation one
has to take the symmetric product when going to the noncommutative
realm: $AB \rightarrow \frac{1}{2}\{A,B\}_{\star}$. Compatibility with
gauge transformations implies that only the sum of the conformal
transformations of gauge field $\hat{A}$ and $\theta$ has a meaning. A
covariant splitting of this sum allows a $\theta$-expansion into a
commutative gauge theory.

\subsection{Conformal transformations of the noncommutative gauge field}
\label{sec51}

We generalize the (rigid) conformal transformations
(\ref{TA})--(\ref{DA}) to noncommutative Yang-Mills theory, i.e.\ a
gauge theory for the field $\hat{A}_\mu$ transforming according to
(\ref{Ahat}):
\begin{align}
W_{\hat{A};\tau}^T &:= 
\int d^4x\,\mathrm{tr}\Big( \partial_\tau \hat{A}_\mu \,
\frac{\delta}{\delta \hat{A}_\mu} \Big) ~,
\label{WTTA}
\\
W_{\hat{A};\alpha\beta}^R &:= 
\int d^4x\,\mathrm{tr}\Big( 
\Big(\frac{1}{2} \big\{ x_\alpha, 
\partial_\beta \hat{A}_\mu \big\}_\star 
- \frac{1}{2} \big\{ x_\beta, 
\partial_\alpha \hat{A}_\mu \big\}_\star 
+ g_{\mu\alpha} \hat{A}_\beta
- g_{\mu\beta} \hat{A}_\alpha \Big)
\,\frac{\delta}{\delta \hat{A}_\mu} \Big) ~,
\label{WTRA}
\\
W_{\hat{A}}^D &:= 
\int d^4x\,\mathrm{tr}\Big( 
\Big(\frac{1}{2} \big\{ x^\delta, 
\partial_\delta \hat{A}_\mu \big\}_\star 
+ \hat{A}_\mu \Big) \,\frac{\delta}{\delta \hat{A}_\mu} \Big) ~,
\label{WTDA}
\end{align}
where $\big\{ U,V\big\}_\star := U \star V + V \star U$ is the
$\star$-anticommutator. It is important to take the symmetric product
in the ``quantization'' $x_\alpha \partial_\beta A_\mu \mapsto
\tfrac{1}{2} \{x_\alpha,\partial_\beta \hat{A}_\mu\}_\star\,$. Let us
introduce the convenient abbreviation $W^?_{\hat{A}}$ standing for one
of the operators $\{W^T_{\hat{A};\tau}, W^R_{\hat{A};\alpha\beta},
W^D_{\hat{A}}\}$ and similarly for $W^?_\theta$ in
(\ref{TT})--(\ref{DT}). 

Applying $W_{\hat{A};\alpha\beta}^R$ to the noncommutative
Yang-Mills field strength $\hat{F}_{\mu\nu} = \partial_\mu
\hat{A}_\nu - \partial_\nu \hat{A}_\mu -
\mathrm{i}[\hat{A}_\mu,\hat{A}_\nu]_\star$ one obtains
\begin{align}
W^R_{\hat{A};\alpha\beta} \hat{F}_{\mu\nu} 
&= 
\tfrac{1}{2} \{x_\alpha,\partial_\beta \hat{F}_{\mu\nu}\}_\star 
-  \tfrac{1}{2} \{x_\beta,\partial_\alpha \hat{F}_{\mu\nu}\}_\star 
+ g_{\mu\alpha} \hat{F}_{\beta\nu} - g_{\mu\beta} \hat{F}_{\alpha\nu} 
+ g_{\nu\alpha} \hat{F}_{\mu\beta} - g_{\nu\beta} \hat{F}_{\mu\alpha} 
\nonumber
\\*
& 
- \tfrac{1}{2} \theta_\alpha^{~\rho}
\{\partial_\rho \hat{A}_\mu,\partial_\beta \hat{A}_\nu\}_\star 
+ \tfrac{1}{2} \theta_\beta^{~\rho}
\{\partial_\rho \hat{A}_\mu,\partial_\alpha \hat{A}_\nu\}_\star 
\nonumber
\\*
&
+ \tfrac{1}{2} \theta_\alpha^{~\rho}
\{\partial_\rho \hat{A}_\nu,\partial_\beta \hat{A}_\mu\}_\star 
- \tfrac{1}{2} \theta_\beta^{~\rho}
\{\partial_\rho \hat{A}_\nu,\partial_\alpha \hat{A}_\mu\}_\star ~,
\label{WTF} 
\end{align}
which is not the expected Lorentz transformation of the field
strength. However, we must also take the $\theta$-transformation
(\ref{TT})--(\ref{DT}) into account, which acts on the $\star$-product
in the $\hat{A}$-bilinear part of $\hat{F}_{\mu\nu}$. Using the
differentiation rule for the $\star$-product
\begin{align}
W^?_{\theta}(U \star V) 
=&\left(W^?_{\theta} U \right) \star V 
+ U \star \left(W^?_{\theta} V\right) 
+
 \frac{\mathrm{i}}{2}\left(W^?_{\theta}\theta^{\mu\nu}\right)
(\partial_\mu U) \star (\partial_\nu V)~,
\label{true} 
\end{align}
which is a consequence of (\ref{ny-1}) and (\ref{dt}), together with 
\begin{align}
  W^?_{\theta}\hat{A}_\mu =0~,
\label{naiv}
\end{align}
one finds that $W^R_{\theta;\alpha\beta} \hat{F}_{\mu\nu}$
cancels exactly the last two lines in (\ref{WTF}):
\begin{align}
(W^R_{\hat{A};\alpha\beta} +W^R_{\theta;\alpha\beta} )
\hat{F}_{\mu\nu} 
&= \tfrac{1}{2} \{x_\alpha,\partial_\beta \hat{F}_{\mu\nu}\}_\star 
-  \tfrac{1}{2} \{x_\beta,\partial_\alpha \hat{F}_{\mu\nu}\}_\star 
\nonumber
\\
& + g_{\mu\alpha} \hat{F}_{\beta\nu} - g_{\mu\beta} \hat{F}_{\alpha\nu} 
+ g_{\nu\alpha} \hat{F}_{\mu\beta} - g_{\nu\beta} \hat{F}_{\mu\alpha} ~.
\label{RF}
\end{align}
In the same way one finds 
\begin{align}
(W^D_{\hat{A}} +W^D_\theta )
\hat{F}_{\mu\nu} &= \tfrac{1}{2} \big\{ x^\delta , 
\partial_\delta \hat{F}_{\mu\nu}
\big\}_\star + 2 \hat{F}_{\mu\nu}~.
\end{align}
It is then easy to verify that the noncommutative Yang-Mills (NCYM) 
action
\begin{equation}
\hat{\Sigma} = -\frac{1}{4g^2} \int d^4x \, \mathrm{tr}(\hat{F}^{\mu\nu} 
\star \hat{F}_{\mu\nu})
\label{NCYM}
\end{equation}
is invariant under noncommutative translations, rotations
and dilatations\footnote{In \cite{Gerhold:2000ik} we have shown that
  an identity like $W^D_\phi \hat{\Sigma}- 2 \theta^{\mu\nu}
  (\partial \hat{\Sigma}/\partial \theta^{\mu\nu}) = 0$ exists for
  dilatation in the case of noncommutative $\phi^4$ theory.}:
\begin{align}
W^T_{\hat{A}+\theta;\tau} \hat{\Sigma} &= 0~, & 
W^R_{\hat{A}+\theta;\alpha\beta} \hat{\Sigma} &= 0~, & 
W^D_{\hat{A}+\theta} \hat{\Sigma} &= 0~, 
\end{align}
with the general notation
\begin{equation}
W^?_{A;C} + W^?_{B;C} =W^?_{A+B;C}~.
\end{equation}

Computing the various commutators between $W^?_{\hat{A}}$ given in
(\ref{WTTA})--(\ref{WTDA}) one convinces oneself that the
noncommutative gauge field $\hat{A}_\mu$ forms an irreducible
representation of the conformal Lie algebra (\ref{cr}). For
convenience we list these commutators (for $W^?_{\hat{A}+\theta}$,
which makes no difference to $W^?_{\hat{A}}$ when applied to
$\hat{A}_\mu$) below in (\ref{CC}). It is remarkable that the
conformal group remains the same and should not be deformed when
passing from a commutative space to a noncommutative one whereas the
gauge groups are very different in both cases. This shows that the
fundamentals of quantum field theory---Lorentz covariance, locality,
unitarity---have good chances to survive in the noncommutative
framework.

In particular, the Wigner theorem \cite{Bargmann:1948ck} that a field
is classified by mass and spin holds. The conformal Lie
algebra is of rank 2, hence its irreducible representations $\rho$ are
(in nondegenerate cases) classified by two Casimir operators,
\begin{align}
  \label{Casimir}
m^2 = -g^{\tau\sigma} \rho(P_\tau) \rho(P_\sigma) ~,\qquad 
s(s+1) m^2 = - g_{\mu\nu} W^{PL;\mu} W^{PL;\nu}~, 
\end{align}
where 
\begin{align}
  \label{PL}
W^{PL;\mu} = - \frac{1}{2} \epsilon^{\mu\tau\alpha\beta}
\rho(P_\tau) \rho(M_{\alpha\beta}) 
\end{align}
is the Pauli-Ljubanski vector and $m$ and $s$ mass and spin of the particle,
respectively. In our case where $\rho(?)$ is given by the action of
$W^?_{\hat{A}+\theta}$ on $\hat{A}_\mu$ we find 
\begin{align}
m^2 \hat{A}_\mu &= 
-\partial^\tau \partial_\tau \hat{A}_\mu ~, &
g_{\rho\sigma} W^{PL;\rho}_{\hat{A}} W^{PL;\sigma}_{\hat{A}} 
\hat{A}_\mu &= 2 (g_{\mu\tau} \partial^\sigma
\partial_\sigma - \partial_\mu \partial_\tau ) \hat{A}^\tau 
+ 0\,  \partial_\mu \partial_\tau \hat{A}^\tau ~,
\end{align}
which means that the transverse components of $\hat{A}_\mu$ have spin
$s=1$ and the longitudinal component spin $s=0$.

\subsection{Compatibility with gauge symmetry}
\label{sec52}

The NCYM action (\ref{NCYM}) is additionally invariant under
noncommutative gauge transformations
\begin{align}
W^G_{\hat{A};\hat{\lambda}} = \int d^4x\,\mathrm{tr} \Big(
\big(\partial_\mu \lambda - \mathrm{i}
\big[\hat{A}_\mu,\hat{\lambda}\big]_\star \big) \,\frac{\delta}{ 
\delta \hat{A}_\mu} \Big)~,
\label{nc-gauge}
\end{align}
where $\hat{\lambda}$ is a possibly $\hat{A}$-dependent gauge
parameter. 
This means that the symmetry algebra of the NCYM action is at 
least\footnote{Renormlizability seems to
  require that the symmetry algebra of the NCYM action is actually
  bigger than $\mathcal{L}$.} given by the Lie algebra 
\begin{align}
\mathcal{L}=\mathcal{G} >\!\!\! \triangleleft\ \mathcal{C}  
\label{semi}
\end{align}
of Ward identity operators, which 
is the semidirect product of the Lie algebra $\mathcal{G}$ of possibly 
field-dependent gauge transformations $W^G_{\hat{A};\hat{\lambda}}$ 
with the Lie algebra $\mathcal{C}$ of rigid conformal transformations
$W^{\{T,R,D\}}_{\hat{A}+\theta}$. The commutator relations of
$\mathcal{L}$ are computed to 
\begin{align}
[W^G_{\hat{A};\hat{\lambda}_1},W^G_{\hat{A};\hat{\lambda}_2}] 
&= -\mathrm{i} W^G_{\hat{A};
[\hat{\lambda}_1, \hat{\lambda}_2]_\star
+ \mathrm{i} W^G_{\hat{A};\hat{\lambda}_1}\hat{\lambda}_2
-\mathrm{i} W^G_{\hat{A};\hat{\lambda}_2} \hat{\lambda}_1 } ~,
\label{GG}
\\*[1ex]
[W^T_{\hat{A}+\theta;\tau} ,W^G_{\hat{A};\hat{\lambda}}]
&= W^G_{\hat{A};- \partial_\tau \hat{\lambda} 
+ W^T_{\hat{A}+\theta;\tau} \hat{\lambda}}~,
\nonumber
\\*
[W^R_{\hat{A}+\theta;\alpha\beta},W^G_{\hat{A};\hat{\lambda}}] 
&= W^G_{\hat{A};
-\frac{1}{2}\{x_\alpha,\partial_\beta \hat{\lambda} \}_\star 
+\frac{1}{2}\{x_\beta,\partial_\alpha \hat{\lambda} \}_\star 
+W^R_{\hat{A}+\theta;\alpha\beta} \hat{\lambda}}~,
\nonumber
\\*
[W^D_{\hat{A}+\theta}, W^G_{\hat{A};\hat{\lambda}}] 
&= W^G_{\hat{A};
-\frac{1}{2}\{x^\delta,\partial_\delta \hat{\lambda} \}_\star 
+W^D_{\hat{A}+\theta} \hat{\lambda}}~,
\label{CG}
\\[1ex]
[W^T_{\hat{A}+\theta;\tau},W^T_{\hat{A}+\theta;\sigma}] &= 0~,
\nonumber
\\{}
[W^T_{\hat{A}+\theta;\tau},W^R_{\hat{A}+\theta;\alpha\beta}] &=
g_{\tau\beta} W^T_{\hat{A}+\theta;\alpha}
-g_{\tau\alpha} W^T_{\hat{A}+\theta;\beta} ~,
\nonumber
\\{}
[W^T_{\hat{A}+\theta;\tau},W^D_{\hat{A}+\theta}] &=
-W^T_{\hat{A}+\theta;\tau} ~,
\nonumber
\\
[W^R_{\hat{A}+\theta;\alpha\beta},W^R_{\hat{A}+\theta;\gamma\delta}] &=
g_{\alpha\gamma} W^R_{\hat{A}+\theta;\beta\delta}
-g_{\beta\gamma} W^R_{\hat{A}+\theta;\alpha\delta}
-g_{\alpha\delta} W^R_{\hat{A}+\theta;\beta\gamma}
+g_{\beta\delta} W^R_{\hat{A}+\theta;\alpha\gamma}~,
\nonumber
\\*
[W^R_{\hat{A}+\theta;\alpha\beta},W^D_{\hat{A}+\theta}] &=0~,
\nonumber
\\*
[W^D_{\hat{A}+\theta},W^D_{\hat{A}+\theta}] &=0~.
\label{CC}
\end{align}
It is crucial to use the sum of the individual transformations
$W^{\{R,D\}}_{\hat{A}}$ and $W^{\{R,D\}}_\theta$ because the
individual commutators do not preserve the Lie algebra $\mathcal{L}$:
\begin{align}
[W^G_{\hat{A};\hat{\lambda}},W^R_{\theta;\alpha\beta}] \hat{A}_\mu &= 
W^G_{\hat{A};- W^R_{\theta;\alpha\beta} \hat{\lambda} } \hat{A}_\mu 
-\tfrac{1}{2} \theta_\beta^{~\rho} \{\partial_\alpha\hat{A}_\mu,
\partial_\rho \hat{\lambda}\}_\star 
+\tfrac{1}{2} \theta_\alpha^{~\rho} \{\partial_\beta\hat{A}_\mu,
\partial_\rho \hat{\lambda}\}_\star 
\nonumber
\\
&
+\tfrac{1}{2} \theta_\beta^{~\rho} \{\partial_\rho\hat{A}_\mu,
\partial_\alpha \hat{\lambda}\}_\star 
-\tfrac{1}{2} \theta_\alpha^{~\rho} \{\partial_\rho\hat{A}_\mu,
\partial_\beta \hat{\lambda}\}_\star ~,
\nonumber
\\{}
[W^G_{\hat{A};\hat{\lambda}},W^D_{\theta}] \hat{A}_\mu &= 
W^G_{\hat{A};-W^D_{\theta}\hat{\lambda}}\hat{A}_\mu 
+ \theta^{\delta\rho} 
\{\partial_\delta \hat{A}_\mu,\partial_\rho \hat{\lambda}\}_\star~.
\end{align}

\subsection{Gauge covariance, covariant representation and
Seiberg-Witten differential equation}
\label{sec53}

One may ask (the reason is given below) whether there exists a
`rotation' in $(\hat{A},\theta)$ space so that the `rotated
fields' preserve individually the mixed commutators (\ref{CG}). To be
concrete, what we look for is a splitting
\begin{align}
W^?_{\hat{A}+\theta} \equiv 
W^?_{\hat{A}} + W^?_{\theta} &= \tilde{W}^?_{\hat{A}} +
\tilde{W}^?_{\theta} ~,
\label{split}
\\{}
[\tilde{W}^?_{\hat{A}},W^G_{\hat{A};\hat{\lambda}}] &=
W^G_{\hat{A};\hat{\lambda}^?_{\hat{A}}} ~,\qquad\qquad
[\tilde{W}^?_{\theta},W^G_{\hat{A};\hat{\lambda}}] =
W^G_{\hat{A};\hat{\lambda}^?_\theta} ~,
\label{covariance}
\end{align}
for appropriate field-dependent gauge parameters
$\hat{\lambda}^?_{\hat{A}}$ and $\hat{\lambda}^?_\theta$. Because of
(\ref{CG}), each of the two relations in (\ref{covariance}) is of
course the consequence of the other relation.  Furthermore, we impose
the condition that the splitting should be universal in the sense 
$\tilde{W}^?_{\theta} = W^?_{\theta} (\theta^{\rho\sigma}) 
\frac{d}{d \theta^{\rho\sigma}}$:
\begin{align}
\tilde{W}^?_{\hat{A}} &= W^?_{\hat{A}} 
- W^?_\theta(\theta^{\rho\sigma}) \int d^4x\,\mathrm{tr}\Big( 
\frac{d \hat{A}_\mu}{d \theta^{\rho\sigma}} 
\frac{\delta}{\delta \hat{A}_\mu}\Big)~, 
\nonumber
\\*
\tilde{W}^?_{\theta} &= W^?_{\theta} 
+ W^?_\theta(\theta^{\rho\sigma}) \int d^4x\,\mathrm{tr}\Big( 
\frac{d \hat{A}_\mu}{d \theta^{\rho\sigma}} 
\frac{\delta}{\delta \hat{A}_\mu}\Big) \equiv
W^?_{\theta} (\theta^{\rho\sigma}) 
\frac{d}{d \theta^{\rho\sigma}}~.
\label{WTT}
\end{align}
The notation $\frac{d \hat{A}_\mu}{d \theta^{\rho\sigma}}$ is for the
time being just a symbol for a field-dependent quantity with three
Lorentz indices and power-counting dimension 3. Inserted into
(\ref{covariance}) one gets the \emph{equivalent conditions}
\begin{align}
-\mathrm{i} \big[\tilde{W}^?_{\hat{A}} \hat{A}_\mu, \hat{\lambda} \big]_\star 
- W^G_{\hat{A};\hat{\lambda}} \big(\tilde{W}^?_{\hat{A}}(\hat{A}_\mu) \big)
&= \hat{D}_\mu\big(\hat{\lambda}^?_{\hat{A}} - 
\tilde{W}^?_{\hat{A}} (\hat{\lambda})\big)~,
\label{cA}
\\*
W^?_\theta(\theta^{\rho\sigma}) \Big(
-\mathrm{i} \Big[\frac{d \hat{A}_\mu}{d \theta^{\rho\sigma}}, 
\hat{\lambda} \Big]_\star 
+ \frac{1}{2} \big\{\partial_\rho \hat{A}_\mu, \partial_\sigma
\hat{\lambda} \big\}_\star - W^G_{\hat{A};\hat{\lambda}} \Big(
\frac{d \hat{A}_\mu}{d \theta^{\rho\sigma}}\Big)\Big) 
&= \hat{D}_\mu\big(\hat{\lambda}^?_\theta- 
\tilde{W}^?_\theta (\hat{\lambda})\big)~.
\label{cT}
\end{align}
Whereas (\ref{cT}) cannot be solved without prior knowledge of the
result\footnote{One can make of course an ansatz for $\frac{d
    \hat{A}_\mu}{d \theta^{\rho\sigma}}$ with free coefficients
  to be determined by (\ref{cT}).}, we can trivially solve (\ref{cA})
by a covariance ansatz:
\begin{align}
\tilde{W}^T_{\hat{A};\tau} &= W^G_{\hat{A};\hat{\lambda}^T_\tau} 
+ \int d^4x\,\mathrm{tr}\Big(
\hat{F}_{\tau\mu} \frac{\delta}{\delta \hat{A}_\mu} \Big) ~,
\label{TTA} 
\\*
\tilde{W}^R_{\hat{A};\alpha\beta} &= 
W^G_{\hat{A};\hat{\lambda}^R_{\alpha\beta}} 
+\int d^4x\,\mathrm{tr}\Big(\Big(
\frac{1}{2} \{\hat{X}_\alpha,\hat{F}_{\beta\mu}\}_\star 
-\frac{1}{2} \{\hat{X}_\beta,\hat{F}_{\alpha\mu}\}_\star 
- W^R_{\theta;\alpha\beta} (\theta^{\rho\sigma})
\hat{\Omega}_{\rho\sigma\mu} \Big)
\frac{\delta}{\delta \hat{A}_\mu} \Big)~,
\label{TRA}
\\*
\tilde{W}^D_{\hat{A}} &= W^G_{\hat{A};\hat{\lambda}^D} 
+ \int d^4x\,\mathrm{tr}\Big(\Big(
\frac{1}{2} \{\hat{X}^\delta,\hat{F}_{\delta\mu}\}_\star 
- W^D_{\theta} (\theta^{\rho\sigma})
\hat{\Omega}_{\rho\sigma\mu} \Big)
\frac{\delta}{\delta \hat{A}_\mu} \Big)~,
\label{TDA}
\end{align}
where $\hat{X}^\mu=x^\mu+\theta^{\mu\nu}\hat{A}_\nu$ are the covariant
coordinates \cite{Cornalba:2000ah,Madore:2000en} and
$\hat{\Omega}_{\rho\sigma\mu}$ is a polynomial in the covariant
quantities $\theta,\hat{X},\hat{F},\hat{D}\dots\hat{D} \hat{F}$ which
is antisymmetric in $\rho,\sigma$ and of power-counting dimension $3$.
For physical reasons (e.g.\ quantization) an $\hat{X}$-dependence of
$\hat{\Omega}_{\rho\sigma\mu}$ should be excluded. We denote
(\ref{TTA})--(\ref{TDA}) as \emph{covariant transformations} of the
noncommutative gauge field $\hat{A}$, because these transformations
reduce in the commutative case to the `gauge-covariant conformal
transformations' of Jackiw \cite{Jackiw:1978ar,Jackiw:1980}.

It follows from (\ref{semi}) and (\ref{split}) that
$\tilde{W}^?_\theta$ and thus $\frac{d \hat{A}_\mu}{d
  \theta^{\rho\sigma}}$ are (up to a gauge transformation) precisely
the missing piece to complete (\ref{TRA}) and (\ref{TDA}) to an
invariance of the action,
\begin{align}
(\tilde{W}^R_{\hat{A};\alpha\beta} +\tilde{W}^R_{\theta;\alpha\beta}) 
\hat{\Sigma}=0 ~,\qquad 
(\tilde{W}^D_{\hat{A};\alpha\beta} +\tilde{W}^D_{\theta;\alpha\beta}) 
\hat{\Sigma}=0 ~.
\label{sumr}
\end{align}
Applying (\ref{TTA})--(\ref{TDA}) to the NCYM action (\ref{NCYM}) we
obtain for $\hat{\Omega}_{\rho\sigma\mu}=0$
\begin{align}
\tilde{W}^T_{\hat{A};\tau} \hat{\Sigma} &=0~,
\label{ST}
\\*
\tilde{W}^R_{\hat{A};\alpha\beta} \hat{\Sigma} &= 
\frac{1}{g^2} \int d^4x \,\mathrm{tr} \Big( 
\theta_{\alpha\rho} \hat{F}^{\rho\sigma} \star \hat{T}_{\beta\sigma} 
- \theta_{\beta\rho} \hat{F}^{\rho\sigma} \star \hat{T}_{\alpha\sigma} 
\Big)~,
\label{SR}
\\*
\tilde{W}^D_{\hat{A}} \hat{\Sigma} &= 
\frac{1}{g^2} \int d^4x \,\mathrm{tr} \Big( 
\theta^\delta_{~\rho} \hat{F}^{\rho\sigma} \star \hat{T}_{\delta\sigma} 
\Big)~,
\label{SD}
\end{align}
where the quantity 
\begin{align}
  \label{EMT}
\hat{T}_{\mu\nu} = 
\frac{1}{2} \hat{F}_{\mu\rho}\star \hat{F}_\nu^{~\rho}
+\frac{1}{2}\hat{F}_{\nu\rho} \star\hat{F}_\mu^{~\rho}
- \frac{1}{4} g_{\mu\nu} \hat{F}_{\rho\sigma}\star
\hat{F}^{\rho\sigma} 
\end{align}
resembles (but is not) the energy-momentum tensor. The calculation
uses however the symmetry $\hat{T}_{\mu\nu}=\hat{T}_{\nu\mu}$ 
(a consequence of the symmetrical product in (\ref{TRA})) and
tracelessness $g^{\mu\nu} \hat{T}_{\mu\nu}=0$. We give in Appendix
\ref{appa} details of the computation of (\ref{SR}). As we show in
Appendix \ref{appb}, the first (rotational) condition in (\ref{sumr})
has, reinserting $\hat{\Omega}_{\rho\sigma\mu}$, the solution
\begin{align}
\label{SWA}
\frac{d \hat{A}_\mu}{d \theta^{\rho\sigma}} =  
-\frac{1}{8} \big\{ \hat{A}_\rho, \partial_\sigma \hat{A}_\mu +
\hat{F}_{\sigma\mu} \big\}_\star 
+\frac{1}{8} \big\{ \hat{A}_\sigma, \partial_\rho \hat{A}_\mu +
\hat{F}_{\rho\mu} \big\}_\star + \hat{\Omega}_{\rho\sigma\mu}~,
\end{align}
which is also compatible with the second (dilatational) condition in
(\ref{sumr}). The solution (\ref{SWA}) is for 
$\hat{\Omega}_{\rho\sigma\mu}=0$ known as the Seiberg-Witten
differential equation \cite{Seiberg:1999vs}.
It is now straightforward to check (\ref{cT}) for an arbitrary
field-dependent gauge parameter $\hat{\lambda}$. The gauge parameters
in (\ref{WTT}) are 
\begin{align}
\hat{\lambda}^T_\tau &= \hat{A}_\tau ~,
&
\hat{\lambda}^R_{\alpha\beta} &= 
\frac{1}{4}\{2 x_\alpha+\theta_\alpha^{~\rho}\hat{A}_\rho,
\hat{A}_\beta\}_\star
-\frac{1}{4}\{2x_\beta+\theta_\beta^{~\rho}\hat{A}_\rho,
\hat{A}_\alpha\}_\star~, 
&
\hat{\lambda}^D &=\frac{1}{2}\{x^\delta,\hat{A}_\delta\}_\star~.
\label{lambda}
\end{align}

\subsection{$\theta$-expansion of noncommutative gauge
transformations}
\label{sec54}

The meaning of the condition (\ref{covariance}) is easy to
understand: $\tilde{W}^?_\theta$ applied to a
gauge-invariant functional remains gauge-invariant. Because
$\tilde{W}^?_\theta(\theta^{\rho\sigma})$ commutes with
$W^G_{\hat{A};\hat{\lambda}}$, we conclude with the notation 
$\frac{d}{d \theta^{\rho\sigma}} = 
\frac{\partial}{\partial \theta^{\rho\sigma}} + 
\int d^4x \,\mathrm{tr} \big(
\frac{d \hat{A}_\mu }{d \theta^{\rho\sigma}} 
\frac{\delta}{\delta \hat{A}_\mu} \big)$, see (\ref{WTT}), 
 that
\begin{align}
\Big[ \frac{d}{d \theta^{\rho\sigma}},
W^G_{\hat{A};\hat{\lambda}}\Big] 
= W^G_{\hat{A};\hat{\lambda}_{\rho\sigma}(\hat{\lambda})}~,
\label{tG}
\end{align}
where $\hat{\lambda}_{\rho\sigma}(\hat{\lambda})$ is determined by
$\hat{\lambda}$ and the choice $\frac{d \hat{A}_\mu}{d
  \theta^{\rho\sigma}}$. In particular, we conclude from (\ref{tG}) that
\begin{align}
  \frac{d^n \Gamma}{d \theta^{\rho_1\sigma_1} \dots d
    \theta^{\rho_n\sigma_n}} \quad \text{is gauge-invariant if
    $\Gamma$ is gauge-invariant.}
\label{GI}
\end{align}
Given any first-order differential equation $\frac{d
  \hat{A}_\mu}{d \theta^{\rho\sigma}} = \Phi_{\rho\sigma\mu}
[\hat{A},\theta]$ we can express $\hat{A}$ in terms of $\theta$
and the initial value $A$ at $\theta=0$. In the same way, the
first-order differential equation expresses any
(sufficiently regular) functional $\Gamma[\hat{A},\theta]$ in terms of
$\theta$ and the initial value $A$:
\begin{align}
\Gamma[A,\theta] := \sum_{n=0}^\infty \frac{1}{n!}  \,
\theta^{\rho_1\sigma_1} \cdots \theta^{\rho_n\sigma_n} \,
\Big(\frac{d^n \Gamma[\hat{A},\theta]}{
d \theta^{\rho_1\sigma_1} \dots   
d \theta^{\rho_n\sigma_n}}  \Big)_{\theta=0}~.
\label{taylor}
\end{align}
The special choice (\ref{SWA}) of the differential equation has due to
(\ref{GI}) the distinguished property that 
\begin{align}
W^G_{\hat{A};\hat{\lambda}} \big(\Gamma[\hat{A},\theta]\big) = 0 \quad
\Rightarrow \quad 
W^G_{A;\lambda=\hat{\lambda}|_{\theta=0}} 
\Big(\sum_{n=0}^N \frac{1}{n!}  \,
\theta^{\rho_1\sigma_1} \cdots \theta^{\rho_n\sigma_n} \,
\Big(\frac{d^n \Gamma[\hat{A},\theta]}{
d \theta^{\rho_1\sigma_1} \dots   
d \theta^{\rho_n\sigma_n}}  \Big)_{\theta=0} \Big) = 0 ~.
\label{comGI}
\end{align}
In other words, \emph{any approximation up to order $N$ in $\theta$ of
  a noncommutatively gauge-invariant functional
  $\Gamma[\hat{A},\theta]$ is invariant under commutative gauge
  transformations if the $\theta$-evolution is given by $(\ref{SWA})$,
  i.e.\ the solution of $(\ref{covariance})$}. We stress that the
noncommutative conformal transformations (\ref{WTTA})--(\ref{WTDA})
and their commutators (\ref{CG}) with gauge transformations enabled us
to compute the gauge-equivalent $\theta$-expansion of Seiberg and
Witten directly (without an ansatz) via the equivalent but much
simpler solution of (\ref{sumr}) for the trivially obtained covariant
transformations (\ref{WTT}). 

Our condition (\ref{covariance}) is more general than the original
gauge-equivalence requirement \cite{Seiberg:1999vs} by Seiberg and
Witten. To see this we consider the $\theta$-expansion of
$W^G_{\hat{A};\hat{\lambda}} \hat{A}_\mu$ taking (\ref{tG}) into
account, where $\hat{\lambda}$ is allowed to depend on $\hat{A}$. To
demonstrate the relation we consider the term to second order in
$\theta$:
\begin{align*}
&\frac{d^2 W^G_{\hat{A};\hat{\lambda}} \hat{A}_\mu}{
d \theta^{\rho_1\sigma_1} d \theta^{\rho_2\sigma_2}} 
= \frac{d}{d \theta^{\rho_1\sigma_1}} 
\Big(\Big[\frac{d}{d \theta^{\rho_2\sigma_2}} ,
W^G_{\hat{A};\hat{\lambda}} \Big] + 
W^G_{\hat{A};\hat{\lambda}} 
\frac{d}{d \theta^{\rho_2\sigma_2}} \Big) \hat{A}_\mu
\\
& ~~=\Big(
W^G_{\hat{A};\hat{\lambda}_{\rho_1\sigma_1}
(\hat{\lambda}_{\rho_2\sigma_2}(\hat{\lambda}))} 
+ W^G_{\hat{A};\hat{\lambda}_{\rho_2\sigma_2}(\hat{\lambda})}
\frac{d}{d \theta^{\rho_1\sigma_1}} 
+ W^G_{\hat{A};\hat{\lambda}_{\rho_1\sigma_1}(\hat{\lambda})} 
\frac{d}{d \theta^{\rho_2\sigma_2}} 
+ W^G_{\hat{A};\hat{\lambda}} \frac{d^2}{
d \theta^{\rho_1\sigma_1} d \theta^{\rho_2\sigma_2}}
\Big) \hat{A}_\mu~.
\end{align*}
Setting $\theta \to 0$, generalizing it to any order $n$ and inserting
the result into the Taylor expansion (\ref{taylor}) we obtain 
\begin{align}
\big(W^G_{\hat{A};\hat{\lambda}} \hat{A}_\mu\big) [A,\theta]
&= W^G_{A;\lambda[\hat{\lambda};A,\theta]}
\big(\hat{A}_\mu[A,\theta]\big)~,
\label{GEQ}
\\
\lambda[\hat{\lambda};A,\theta] &= \big(\hat{\lambda}\big)_{\theta=0} 
+ \theta^{\rho\sigma}
\big(\hat{\lambda}_{\rho\sigma}(\hat{\lambda})\big)_{\theta=0} 
+ \tfrac{1}{2} \theta^{\rho_1\sigma_1} \theta^{\rho_2\sigma_2}
\big(\hat{\lambda}_{\rho_1\sigma_1}
(\hat{\lambda}_{\rho_2\sigma_2}(\hat{\lambda}))\big)_{\theta=0} 
+ \dots ~.
\nonumber
\end{align}
Eq.\ (\ref{GEQ}) is the original Seiberg-Witten gauge-equivalence
\cite{Seiberg:1999vs} iff
$\big(\hat{\lambda}_{\rho\sigma}(\hat{\lambda})\big)_{\theta=0} =0$.
In other words, our approach via (\ref{covariance})---which leads to
the same $\theta$-expansion as the Seiberg-Witten requirement, see
(\ref{comGI})---is more general.

\subsection{$\theta$-expansion of noncommutative conformal transformations}
\label{sec55}

According to (\ref{taylor}) let us compute the $\theta$-expansion of
the noncommutative conformal transformation of a functional
$\Gamma[\hat{A},\theta]$ approximated up to order $N$ in $\theta$, 
\begin{align}
\sum_{n=0}^N \frac{1}{n!}  \,
\theta^{\rho_1\sigma_1} \cdots \theta^{\rho_n\sigma_n} \,
\Big(\frac{d^n (W^?_{\hat{A}+\theta} \Gamma[\hat{A},\theta])}{
d \theta^{\rho_1\sigma_1} \dots   
d \theta^{\rho_n\sigma_n}}  \Big)_{\theta=0}~. 
\label{W?G}
\end{align}
As a typical example we regard the $n=2$ term in this series, which we
derive by the following procedure. Before putting $\theta=0$ we consider
\begin{align}
T_2^? &:=
\theta^{\rho_1\sigma_1} \theta^{\rho_2\sigma_2} 
\frac{d^2 (W^?_{\hat{A}+\theta} \Gamma[\hat{A};\theta])}{
d \theta^{\rho_1\sigma_1} d \theta^{\rho_2\sigma_2}}
\nonumber
\\
& = \theta^{\rho_1\sigma_1} \frac{d}{d \theta^{\rho_1\sigma_1}} 
\Big( \theta^{\rho_2\sigma_2} \frac{d (
W^?_{\hat{A}+\theta} \Gamma[\hat{A};\theta])}{d
\theta^{\rho_2\sigma_2}} \Big)
- \theta^{\rho_2\sigma_2} \frac{d (
W^?_{\hat{A}+\theta} \Gamma[\hat{A};\theta])}{d 
\theta^{\rho_2\sigma_2}} ~.
\label{T21}
\end{align}
The crucial property we use is the identity
\begin{align}
\Big[W^?_{\hat{A}+\theta}, \theta^{\rho\sigma} 
\frac{d}{d \theta^{\rho\sigma}} \Big] =0~,
\label{nt}
\end{align}
which is valid for a very general class of differential equations. 
See Appendix \ref{appc} for details. Thus,
\begin{align}
T_2^?
&= W^?_{\hat{A}+\theta} \Big(
\theta^{\rho_1\sigma_1} \frac{d}{d \theta^{\rho_1\sigma_1}} \Big( 
\theta^{\rho_2\sigma_2} \frac{d \Gamma[\hat{A};\theta]}{
d \theta^{\rho_2\sigma_2}} \Big)
- \theta^{\rho_2\sigma_2} \frac{d \Gamma[\hat{A};\theta]}{
d \theta^{\rho_2\sigma_2}} \Big)
= W^?_{\hat{A}+\theta} \Big(
\theta^{\rho_1\sigma_1} 
\theta^{\rho_2\sigma_2} \frac{d^2 \Gamma[\hat{A};\theta]}{
d \theta^{\rho_1\sigma_1}
d \theta^{\rho_2\sigma_2}} \Big)
\nonumber
\\*
&= 
\big(W^?_{\hat{A}+\theta}(\theta^{\rho_1\sigma_1})
\theta^{\rho_2\sigma_2} 
+ \theta^{\rho_1\sigma_1}  
W^?_{\hat{A}+\theta}(\theta^{\rho_2\sigma_2})\big)
\frac{d^2 \Gamma[\hat{A};\theta]}{
d \theta^{\rho_1\sigma_1}
d \theta^{\rho_2\sigma_2}} \Big)
+ \theta^{\rho_1\sigma_1} 
\theta^{\rho_2\sigma_2} 
W^?_{\hat{A}+\theta} \Big(\frac{d^2 \Gamma[\hat{A};\theta]}{
d \theta^{\rho_1\sigma_1}
d \theta^{\rho_2\sigma_2}}\Big)
\nonumber
\\*
&= 
\theta^{\rho_1\sigma_1} 
\theta^{\rho_2\sigma_2} \Big(
\frac{\partial W^?_{\theta}(\theta^{\rho\sigma})}{
\partial \theta^{\rho_1\sigma_1}} 
\frac{d^2 \Gamma[\hat{A};\theta]}{
d \theta^{\rho\sigma}
d \theta^{\rho_2\sigma_2}} 
+
\frac{\partial W^?_{\theta}(\theta^{\rho\sigma})}{
\partial \theta^{\rho_2\sigma_2}} 
\frac{d^2 \Gamma[\hat{A};\theta]}{
d \theta^{\rho_1\sigma_1}
d \theta^{\rho\sigma}} 
+
W^?_{\hat{A}+\theta} \Big(\frac{d^2 \Gamma[\hat{A};\theta]}{
d \theta^{\rho_1\sigma_1}
d \theta^{\rho_2\sigma_2}}\Big)\Big)~,
\label{T22}
\end{align}
using the linearity of $W^?_\theta(\theta^{\rho\sigma})$ in
$\theta$. We can now omit the leading factors of $\theta$ from $T_2^?$
in (\ref{T21}) and (\ref{T22}), generalize it to any order $n$ and put
$\theta=0$:
\begin{align}
  \Big(\frac{d^n (W^?_{\hat{A}+\theta} \Gamma[\hat{A},\theta])}{
d \theta^{\rho_1\sigma_1} \dots   
d \theta^{\rho_n\sigma_n}}  \Big)_{\theta=0}
&= \sum_{i=1}^n \frac{\partial W^?_{\theta}(\theta^{\rho\sigma})}{
\partial \theta^{\rho_i\sigma_i}} 
\Big(\frac{d^n \Gamma[\hat{A};\theta]}{
d \theta^{\rho_2\sigma_1}\dots
d \theta^{\rho_{i-1}\sigma_{i-1}}
d \theta^{\rho\sigma} d \theta^{\rho_{i+1}\sigma_{i+1}}
\dots d \theta^{\rho_n\sigma_n}} \Big)_{\theta=0}
\nonumber
\\
&+
W^?_{A} \Big(\frac{d^n \Gamma[\hat{A};\theta]}{
d \theta^{\rho_1\sigma_1} \dots 
d \theta^{\rho_n\sigma_n}}\Big)_{\theta=0}~.
\end{align}
Note that from $W^?_{\hat{A}+\theta}$ at $\theta=0$ 
there survives only the
commutative conformal transformation $W^?_A$ defined in
(\ref{TA})--(\ref{DA}). Inserted into (\ref{W?G}) we get the final
result
\begin{align}
  \sum_{n=0}^N \frac{1}{n!}  \,
\theta^{\rho_1\sigma_1} \cdots \theta^{\rho_n\sigma_n} \,&
\Big(\frac{d^n (W^?_{\hat{A}+\theta} \Gamma[\hat{A},\theta])}{
d \theta^{\rho_1\sigma_1} \dots   
d \theta^{\rho_n\sigma_n}}  \Big)_{\theta=0} 
\nonumber
\\
&= W_{A+\theta} \Big(  \sum_{n=0}^N \frac{1}{n!}  \,
\theta^{\rho_1\sigma_1} \cdots \theta^{\rho_n\sigma_n} \,
\Big(\frac{d^n \Gamma[\hat{A},\theta]}{
d \theta^{\rho_1\sigma_1} \dots   
d \theta^{\rho_n\sigma_n}}  \Big)_{\theta=0} \Big)~.
\end{align}
This result can be formulated as
\begin{thm}
Acting with the noncommutative conformal transformations
(translation, rotation, dilatation) on action
functionals $\Gamma[\hat{A},\theta]$ and applying the Seiberg-Witten
map is \underline{identical} to the action of the commutative 
translation, rotation and dilatation operations, respectively, 
on $\Gamma[\hat{A}[A,\theta],\theta]$. 
\end{thm}
The result means that with the noncommutative conformal symmetries
there are---after Seiberg-Witten map---no further symmetries
associated than the standard commutative conformal symmetries. Thus,
the noncommutative conformal symmetries do not give any hints for the
renormalization of noncommutative Yang-Mills theories.

\section{Quantization}
\label{sec7}

Passing from a classical action with gauge symmetry to quantum field
theory one must introduce gauge-fixing terms to the action in order to
define the propagator. Here we repeat this construction for the
noncommutative Yang-Mills theory. 

The NCYM theory is enlarged by the
fields $\hat{c}, \hat{\bar{c}},\hat{B}$ which transform according to
the following representation of (\ref{cr}):
\begin{align}
W^T_{\hat{A}+\hat{c}+\smallhatbarc+\hat{B}+\theta;\tau} 
&= W^T_{\hat{A}+\theta;\tau} + \int d^4x \,\mathrm{tr} \Big(
\partial_\tau \hat{c}\,\frac{\delta}{\delta \hat{c}} 
+\partial_\tau \hat{\bar{c}}\,\frac{\delta}{\delta \hat{\bar{c}}} 
+\partial_\tau \hat{B}\,\frac{\delta}{\delta \hat{B}} \Big) ~,
\label{WTG}
\\
W^R_{\hat{A}+\hat{c}+\smallhatbarc+\hat{B}+\theta;\alpha\beta} 
&= W^R_{\hat{A}+\theta;\alpha\beta} + 
\int d^4x \,\mathrm{tr} \Big(
\Big(\frac{1}{2} \big\{ x_\alpha ,\partial_\beta \hat{c} \big\}_\star 
- \frac{1}{2} \big\{ x_\beta ,\partial_\alpha \hat{c} \big\}_\star\Big) 
\frac{\delta}{\delta \hat{c}} 
\nonumber
\\*
& \hspace*{10em} 
+\Big(\frac{1}{2} \big\{x_\alpha , \partial_\beta \hat{\bar{c}} \big\}_\star 
-\frac{1}{2} \big\{x_\beta , \partial_\alpha \hat{\bar{c}} \big\}_\star \Big)
\frac{\delta}{\delta \hat{\bar{c}}} 
\nonumber
\\*
& \hspace*{10em} 
+ \Big(\frac{1}{2} \big\{x_\alpha, \partial_\beta \hat{B}\big\}_\star 
- \frac{1}{2} \big\{x_\beta, \partial_\alpha \hat{B}\big\}_\star \Big)
\frac{\delta}{\delta \hat{B}} \Big) ~,
\label{WRG}
\\
W^D_{\hat{A}+\hat{c}+\smallhatbarc + \hat{B}+\theta} 
&= W^D_{\hat{A}+\theta} +  
\int d^4x \,\mathrm{tr} \Big(
\frac{1}{2} \big\{ x^\delta ,\partial_\delta \hat{c} \big\}_\star \,
\frac{\delta}{\delta \hat{c}} 
+\Big(\frac{1}{2} \big\{x^\delta , \partial_\delta \hat{\bar{c}}
\big\}_\star + 2 \hat{\bar{c}} \Big)
\nonumber
\\
& \hspace*{10em} 
+ \Big(\frac{1}{2} \big\{x^\delta, \partial_\delta \hat{B}\big\}_\star 
+ 2 B \Big)
\frac{\delta}{\delta \hat{B}} \Big) ~.
\label{WDG}
\end{align}
The noncommutative BRST transformations are given by
\begin{align}
\hat{s} \hat{A}_\mu &= \hat{D}_\mu \hat{c} ~, &
\hat{s} \hat{c} &= - \mathrm{i} c \star c ~, &
\hat{s} \hat{\bar{c}} &= \hat{B} ~, &
\hat{s} \hat{B} &= 0 ~. 
\end{align}
It is then not difficult to verify that the standard gauge-fixing
action
\begin{align}
  \hat{\Sigma}_{gf} = \int d^4x\,\mathrm{tr}\Big( \hat{s} \Big[ 
\hat{\bar{c}} \star \Big(\partial^\mu \hat{A}_\mu + \frac{\alpha}{2}
\hat{B}\Big) \Big] \Big)
\label{gf}
\end{align}
is conformally invariant:
\begin{align}
 W^T_{\hat{A}+\hat{c}+\smallhatbarc +\hat{B}+\theta;\tau} 
\hat{\Sigma}_{gf} &=0~, &
 W^R_{\hat{A}+\hat{c}+\smallhatbarc +\hat{B}+\theta;\alpha\beta} 
\hat{\Sigma}_{gf} &=0~, &
 W^D_{\hat{A}+\hat{c}+\smallhatbarc +\hat{B}+\theta}  
\hat{\Sigma}_{gf} &=0~.
\end{align}
Loop calculations based on $\hat{\Sigma}+\hat{\Sigma}_{gf}$ in
(\ref{NCYM}) and (\ref{gf}) suffer from infrared divergences
\cite{Matusis:2000jf}.

To circumvent the IR-problem one can however use the
$\theta$-expansion of the NCYM action leading to a gauge field theory on
commutative space-time coupled to an external field $\theta$. This
action is quantized according to the analogous
formulae as above, omitting everywhere the hat
symbolizing noncommutative objects and replacing the $\star$-product by
the ordinary product. This approach was used in \cite{Bichl:2001nf} to
compute the one-loop photon selfenergy in $\theta$-expanded Maxwell
theory and in \cite{Bichl:2001cq} to show renormalizability of the 
photon selfenergy to all orders in $\hbar$ and $\theta$.

\section{Summary and outlook}
\label{sec8}

We have established rigid conformal transformations
(\ref{WTTA})--(\ref{WTDA}) for the noncommutative Yang-Mills field
$\hat{A}$. Our results related to these transformations can be
summarized as follows.

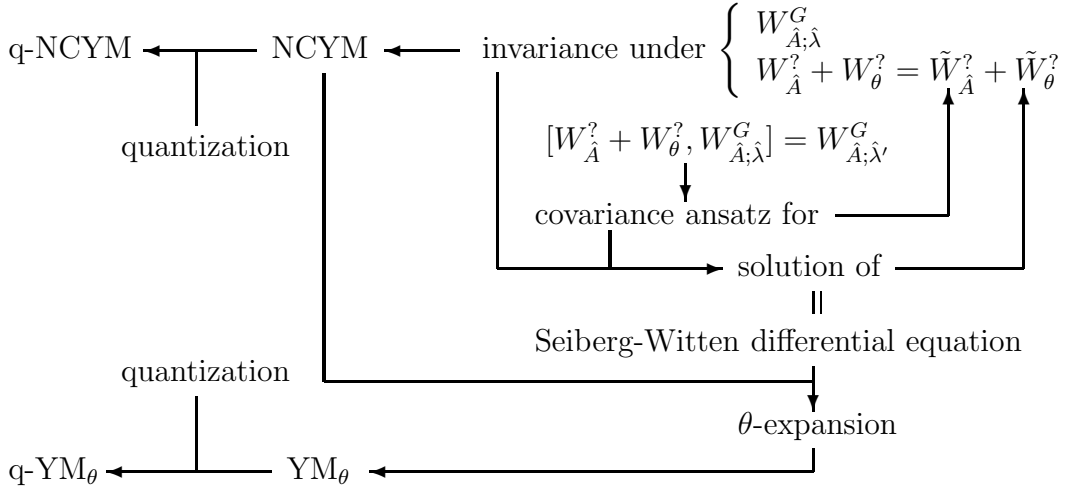
\begin{picture}(155,70)
\thicklines
\put(45,60){NCYM}
\put(70,61){\vector(-1,0){10}}
\put(73,60){invariance under $\left\{
\begin{array}[c]{l}
W^G_{\hat{A};\hat{\lambda}}\\
W^?_{\hat{A}} + W^?_\theta=
\tilde{W}^?_{\hat{A}} + \tilde{W}^?_\theta
\end{array} \right.$}
\put(80,48){
$[W^?_{\hat{A}} + W^?_\theta,W^G_{\hat{A};\hat{\lambda}}]
= W^G_{\hat{A};\hat{\lambda}'}$}
\put(100,46){\vector(0,-1){5}}
\put(80,38){covariance ansatz for}
\put(120,39){\line(1,0){15}}
\put(135,39){\vector(0,1){17}}
\put(75,58){\line(0,-1){26}}
\put(90,37){\line(0,-1){5}}
\put(75,32){\vector(1,0){30}}
\put(107,31){solution of}
\put(128,32){\line(1,0){17}}
\put(145,32){\vector(0,1){24}}
{\thinlines
\put(117,29){\line(0,-1){3}}
\put(118,29){\line(0,-1){3}}}
\put(80,21){Seiberg-Witten differential equation}
\put(52,58){\line(0,-1){41}}
\put(52,17){\line(1,0){65}}
\put(117,19){\vector(0,-1){6}}
\put(107,10){$\theta$-expansion}
\put(117,8){\line(0,-1){3}}
\put(117,5){\vector(-1,0){59}}
\put(47,4){$\text{YM}_\theta$}
\put(25,47){quantization}
\put(35,51){\line(0,1){10}}
\put(25,17){quantization}
\put(35,15){\line(0,-1){10}}
\put(45,5){\vector(-1,0){22}}
\put(10,4){q-$\text{YM}_\theta$}
\put(43,61){\vector(-1,0){15}}
\put(10,60){q-NCYM}
\end{picture}

\vskip 2mm
\noindent
The (classical) noncommutative Yang-Mills action (\ref{NCYM}) is
invariant under the Lie algebra $\mathcal{L}$ of gauge transformations
$W^G_{\hat{A};\hat{\lambda}}$ and the sum $W^?_{\hat{A}} + W^?_\theta$
of conformal transformations of $\hat{A}$ and $\theta$. The
commutation relations $[W^?_{\hat{A}} +
W^?_\theta,W^G_{\hat{A};\hat{\lambda}}] =
W^G_{\hat{A};\hat{\lambda}'}$ in $\mathcal{L}$ suggest a
\emph{covariant} splitting $W^?_{\hat{A}} +
W^?_\theta=\tilde{W}^?_{\hat{A}} + \tilde{W}^?_\theta$. The relation
$[\tilde{W}^?_{\hat{A}} , W^G_{\hat{A};\hat{\lambda}}]=
W^G_{\hat{A};\hat{\lambda}''}$ is trivially solved by a covariance
ansatz. Then, the covariant complement $\tilde{W}^?_\theta$ is simply
obtained from invariance of the NCYM action under
$\tilde{W}^?_{\hat{A}} + \tilde{W}^?_\theta$ transformation. The
solution for $\tilde{W}^?_\theta$ is given by the Seiberg-Witten
differential equation (\ref{SWA}). What we have thus achieved is a
more transparent---and less restrictive---derivation of the
Seiberg-Witten differential equation which does not require the usual
ansatz of gauge equivalence.

Interpreting the Seiberg-Witten differential equation as an evolution
equation we can express the noncommutative Yang-Mills field $\hat{A}$
in terms of its initial value $A$. The resulting $\theta$-expansion of
the NCYM action is due to the covariance $[\tilde{W}^?_\theta ,
W^G_{\hat{A};\hat{\lambda}}]= W^G_{\hat{A};\hat{\lambda}'''}$
invariant under \emph{commutative} gauge transformations. Moreover,
noncommutative conformal transformations reduce after
$\theta$-expansion to commutative conformal transformations. In this
way we associate to the NCYM theory a gauge theory $\text{YM}_\theta$
on commutative space-time for a commutative gauge field $A$ coupled to
a translation-invariant external field $\theta$. Both gauge theories
can be quantized by adding appropriate gauge-fixing terms and yield
the two quantum field theories q-NCYM and q-$\text{YM}_\theta$,
respectively.  It is unclear in which sense these two quantum field
theories are equivalent. At least on a perturbative level the quantum
field theories $\text{q-NCYM}$ and $\text{q-YM}_\theta$ are completely
different.

Loop calculations \cite{Matusis:2000jf} and power-counting analysis
\cite{Chepelev:2001hm} for $\text{q-NCYM}$ reveal a new type of
infrared singularities which so far could not be treated.  Loop
calculations \cite{Bichl:2001nf} for $\text{q-YM}_\theta$ are free
of infrared problems but lead apparently to an enormous amount of
ultraviolet singularities.  This is not necessarily a problem. For
instance, all UV-singularities in the photon selfenergy are
\emph{field redefinitions} \cite{Bichl:2001cq} which are possible in
presence of a field $\theta^{\mu\nu}$ of negative power-counting
dimension. For higher $N$-point Green's functions the situation becomes
more and more involved and a renormalization seems to be impossible
without a symmetry for the $\theta$-expanded NCYM-action. We had hoped
in the beginning of the work on this paper that this symmetry searched
for could be the Seiberg-Witten expansion of the noncommutative
conformal symmetries.  As we have seen in Section \ref{sec55} this is
not the case and the complete renormalization of NCYM theory remains
an open problem.

We have proved that the noncommutative gauge field is an irreducible
representation of the \emph{undeformed} conformal Lie algebra. The
noncommutative spin-$\frac{1}{2}$ representations for fermions have
been worked out in \cite{fermion}. This shows that classical concepts
of particles and fields extend without modification to a
noncommutative space-time. We believe this makes life in a
noncommutative world more comfortable.

Of course much work remains to be done. First we have considered a
very special noncommutative geometry of a constant $\theta^{\mu\nu}$.
This assumption should finally be relaxed; at least the treatment of
those non-constant $\theta^{\mu\nu}$ which are Poisson bivectors as in
\cite{Jurco:2001my} seems to be possible. The influence of the
modified concept of locality on causality and unitarity of the
S-matrix must be studied. Previous results
\cite{Seiberg:2000gc,Gomis:2000zz} with different consequences
according to whether the electrical components of $\theta^{\mu\nu}$
are zero must be invariantly formulated in terms of the signs of the
two invariants $\theta^{\mu\nu}\theta_{\mu\nu}$ and
$\epsilon_{\mu\nu\rho\sigma} \theta^{\mu\nu}\theta^{\rho\sigma}$.
Eventually the renormalization puzzle for noncommutative Yang-Mills
theory ought to be solved.

\section*{Acknowledgements}

We would like to thank Roman Jackiw for numerous interesting comments
and for pointing out to us his earlier work on the covariant
representation \cite{Jackiw:1978ar,Jackiw:1980}. We also thank Alan
Kostelecky for clarifying comments on observer and particle Lorentz
transformations.  JMG would like to thank the University of Bonn for
friendly hospitality during a visit.

\begin{appendix}

\renewcommand{\theequation}{\Alph{section}.\arabic{equation}}

\makeatletter
\@addtoreset{equation}{section}
\makeatother

\section{Covariant $\hat{A}$-rotation of the NCYM action} 
\label{appa}

Let us give here the calculations leading to the result
(\ref{SR}). The first input is the $\hat{A}$-variation of the NCYM
action (\ref{NCYM}) 
\begin{align}
\frac{\delta \hat{\Sigma} }{\delta \hat{A}_\mu(x)} 
= \frac{1}{g^2} \big(\hat{D}_\kappa \hat{F}^{\kappa\mu}\big)(x)~.
\label{appa1}
\end{align}
Inserted into (\ref{TRA}), for $\hat{\Omega}_{\rho\sigma\mu}=0$, we obtain
\begin{align}
\tilde{W}^R_{\hat{A};\alpha\beta} \hat{\Sigma} &=
\frac{1}{2g^2} \int d^4 x\, \mathrm{tr} 
\Big( (\hat{X}_\alpha \star 
\hat{F}_{\beta\mu} + \hat{F}_{\beta\mu} \star \hat{X}_\alpha 
- \hat{X}_\beta \star \hat{F}_{\alpha\mu}
- \hat{F}_{\alpha\mu} \star \hat{X}_\beta ) \star 
\hat{D}_\kappa \hat{F}^{\kappa\mu}
\Big)
\nonumber
\\*
&= \frac{1}{2g^2 } \int d^4 x\, \mathrm{tr} \Big( 
\hat{X}_\alpha \star \big(\hat{D}_\kappa \big\{
\hat{F}_{\beta\mu} , \hat{F}^{\kappa\mu}\big\}_\star
- 
\big\{ \hat{D}_\kappa( \hat{F}_{\beta\mu}) , \hat{F}^{\kappa\mu}
\big\}_\star \big)
\nonumber
\\*
& \hspace*{7em} 
-\hat{X}_\beta \star \big(\hat{D}_\kappa \big\{
\hat{F}_{\alpha\mu} , \hat{F}^{\kappa\mu}\big\}_\star
- 
\big\{ \hat{D}_\kappa( \hat{F}_{\alpha\mu}) , \hat{F}^{\kappa\mu}
\big\}_\star \big)
\Big)\,. 
\label{appa2}
\end{align}
Now we use the Bianchi identity $\hat{D}_\alpha \hat{F}_{\beta\gamma} 
+\hat{D}_\beta \hat{F}_{\gamma\alpha} 
+\hat{D}_\gamma \hat{F}_{\alpha\beta}=0$ and the antisymmetry in
$\kappa,\mu$ to rewrite 
\begin{align}
\hat{D}_\kappa (\hat{F}_{\beta\mu}) \star \hat{F}^{\kappa\mu} 
= \frac{1}{2} \hat{D}_\beta (\hat{F}_{\kappa\mu}) \star 
\hat{F}^{\kappa\mu}
\end{align}
and similarly for the other terms in (\ref{appa2}). We then obtain
\begin{align}
\tilde{W}^R_{\hat{A};\alpha\beta} \hat{\Sigma} 
&= \frac{1}{g^2} \int d^4 x\,  \mathrm{tr} \Big( 
\hat{X}_\alpha \star \hat{D}_\kappa \Big(
\frac{1}{2} \big\{\hat{F}_{\beta\mu}, \hat{F}^{\kappa\mu}\big\}_\star
- \frac{1}{8} \delta_\beta^\kappa  
\big\{\hat{F}_{\mu\nu}, \hat{F}^{\mu\nu}\big\}_\star \Big)
\nonumber
\\*
& \hspace*{5em} -
\hat{X}_\beta \star \hat{D}_\kappa \Big(
\frac{1}{2} \big\{\hat{F}_{\alpha\mu}, \hat{F}^{\kappa\mu}\big\}_\star
- \frac{1}{8} \delta_\alpha^\kappa  
\big\{\hat{F}_{\mu\nu}, \hat{F}^{\mu\nu}\big\}_\star \Big)
\Big)
\nonumber
\\*
&= \frac{1}{g^2} \int d^4 x\,  \mathrm{tr} \Big( 
\hat{D}_\kappa\Big( \hat{X}_\alpha \star \hat{T}_\beta^{~\kappa} 
- \hat{X}_\beta \star \hat{T}_\alpha^{~\kappa} \Big)
- \hat{D}_\kappa (\hat{X}_\alpha) \star 
\hat{T}_\beta^{~\kappa} 
+ \hat{D}_\kappa (\hat{X}_\beta) \star \hat{T}_\alpha^{~\kappa} \Big)~,
\label{appa4}
\end{align}
where we have used (\ref{EMT}) and the derivation property of
$\hat{D}_\kappa$. Note that the total derivative $\int
d^4x\,\mathrm{tr}(\hat{D}_\kappa \hat{J}^\kappa_{\alpha\beta})$ 
in (\ref{appa4}) vanishes. The result (\ref{SR}) follows now from 
\begin{align}
\hat{D}_\kappa \hat{X}_\alpha = g_{\alpha\kappa} +
\theta_\alpha^{~\nu} \hat{F}_{\kappa\nu}~,
\end{align}
which is easily derived from the formulae in Section~\ref{sec2}, and
the symmetry $\hat{T}_{\alpha\beta}=\hat{T}_{\beta\alpha}$.

\section{Derivation of the Seiberg-Witten differential equation}
\label{appb}

We first compute the explicit $\theta$-dependence of the
$\star$-product according to the last term in (\ref{true}), 
\begin{align}
\label{expl}
W_{\theta;\alpha\beta}^R \hat{\Sigma} & = 
-\frac{1}{g^2} \int d^4x \,\mathrm{tr} \Big( 
\theta_{\alpha\rho}
\partial^\rho \hat{A}^\sigma \star \Big\{
\frac{1}{2} \partial_\beta \hat{A}_\nu,  \hat{F}_\sigma^{~\nu}
\Big\}_\star 
-
\theta_{\beta\rho} \partial^\rho \hat{A}^\sigma
\star \Big\{\frac{1}{2} \partial_\alpha \hat{A}_\nu , 
\hat{F}_\sigma^{~\nu} \Big\}_\star 
\Big)~.
\end{align}
Then, (\ref{WTT}) and (\ref{appa1}) yield
\begin{align}
\tilde{W}^R_{\theta;\alpha\beta} \hat{\Sigma} &= 
\text{rhs(\ref{expl})} + \frac{1}{g^2} \int d^4x \,\mathrm{tr} \Big(
\big(
\delta^\rho_\alpha \theta_\beta^{~\sigma} 
- \delta^\rho_\beta \theta_\alpha^{~\sigma} 
+ \delta^\sigma_\alpha \theta^\rho_{~\beta} 
- \delta^\sigma_\beta \theta^\rho_{~\alpha} \big) 
\frac{d \hat{A}_\mu}{d \theta^{\rho\sigma}} 
\star \hat{D}_\kappa  \hat{F}^{\kappa\mu} \Big)
\nonumber
\\*
& = 
\text{rhs(\ref{expl})} + \frac{2}{g^2} \int d^4x \,\mathrm{tr} \Big(
\theta_\alpha^{~\sigma} 
\hat{D}_\kappa \Big(\frac{d \hat{A}_\mu}{d \theta^{\beta\sigma}} 
\Big) \star \hat{F}^{\kappa\mu} 
-  \theta_\beta^{~\sigma} 
\hat{D}_\kappa \Big(\frac{d \hat{A}_\mu}{d \theta^{\alpha\sigma}} 
\Big) \star \hat{F}^{\kappa\mu} 
\Big)~,
\label{appb1}
\end{align}
where \text{rhs(\ref{expl})} stands for the right hand side of
(\ref{expl}). Inserting 
(\ref{SR}), (\ref{expl}) and (\ref{appb1}) into the first condition 
(\ref{sumr}) and
splitting the result into the independent parts with coefficients 
$\theta_{\alpha\rho}/g^2$ and $\theta_{\beta\rho}/g^2$ we find for the
first one 
\begin{align}
0 &= \int d^4 x \,\mathrm{tr} \Big( \hat{F}^{\rho\sigma} 
\star \hat{T}_{\beta\sigma} - \frac{1}{2} \partial^\rho \hat{A}^\sigma 
\star \big\{ \partial_\beta \hat{A}_\nu, \hat{F}_\sigma^{~\nu}
\big\}_\star 
+ 2 g^{\rho\sigma} \hat{D}_\kappa \Big(\frac{d \hat{A}_\mu}{
d \theta^{\beta\sigma}} \Big) \star \hat{F}^{\kappa\mu} 
\Big)
\nonumber
\\
& =  \int d^4 x \,\mathrm{tr} \Big( 
- \frac{1}{2} \partial^\rho \hat{A}^\sigma 
\star \big\{ \hat{D}_\nu \hat{A}_\beta , 
\hat{F}_\sigma^{~\nu} \big\}_\star 
- \frac{1}{8} \partial^\rho \hat{A}_\beta
\star \big\{ \hat{F}_{\mu\nu},\hat{F}^{\mu\nu}\big\}_\star 
- \frac{1}{2} \hat{D}^\sigma \hat{A}^\rho \star \big\{
\hat{F}_{\beta\nu}, \hat{F}_\sigma^{~\nu} \big\}_\star 
\nonumber
\\*
& \hspace*{5em} 
+ \frac{1}{8} \hat{D}_\beta \hat{A}^\rho \star \big\{
\hat{F}_{\mu\nu}, \hat{F}^{\mu\nu} \big\}_\star 
+ 2 g^{\rho\sigma} \hat{D}_\kappa \Big(\frac{d \hat{A}_\mu}{
d \theta^{\beta\sigma}} \Big) \star \hat{F}^{\kappa\mu} 
\Big)
\nonumber
\\
& =  \int d^4 x \,\mathrm{tr} \Big( g^{\rho\sigma} \Big(
- \frac{1}{2} \big\{ \partial_\sigma \hat{A}_\mu ,
\hat{D}_\nu \hat{A}_\beta \big\}_\star  
- \frac{1}{2}  \big\{ \hat{D}_\mu \hat{A}_\sigma,
\hat{F}_{\beta\nu} \big\}_\star 
\nonumber
\\*
& \hspace*{7em} 
- \frac{1}{8} \big\{ \hat{F}_{\sigma\beta},
\hat{F}_{\mu\nu} \big\}_\star 
+ 2 \hat{D}_\mu \Big(\frac{d \hat{A}_\nu}{
d \theta^{\beta\sigma}} \Big) \Big) 
\star \hat{F}^{\mu\nu} 
\Big)
\nonumber
\\
& =  \int d^4 x \,\mathrm{tr} \Big( g^{\rho\sigma} \Big(
\frac{1}{4} \big\{ \hat{D}_\mu \hat{A}_\beta ,
\partial_\sigma \hat{A}_\nu +  \hat{F}_{\sigma\nu} 
\big\}_\star  
- \frac{1}{4}  \big\{ \hat{D}_\mu \hat{A}_\sigma,
\partial_\beta \hat{A}_\nu + \hat{F}_{\beta\nu} \big\}_\star 
\nonumber
\\*
& \hspace*{5em} 
- \frac{1}{8} \big\{ \hat{F}_{\sigma\beta},
\hat{F}_{\mu\nu} \big\}_\star 
+ 2 \hat{D}_\mu \Big(\frac{d \hat{A}_\nu}{
d \theta^{\beta\sigma}} \Big) \Big) 
\star \hat{F}^{\mu\nu} 
\Big)~,
\label{appb2}
\end{align}
where we have used several times cyclicity of the trace, the identity
$\hat{F}_{\beta\nu} = \partial_\beta \hat{A}_\nu - \hat{D}_\nu
\hat{A}_\beta$ and the antisymmetry of $\hat{F}_{\mu\nu}$. Now we
consider
\begin{align}
&\int d^4 x \,\mathrm{tr} \Big(
\big\{  \hat{A}_\beta ,
\hat{D}_\mu (\partial_\sigma \hat{A}_\nu +  \hat{F}_{\sigma\nu}) 
\big\}_\star \star \hat{F}^{\mu\nu} \Big)
= 
\int d^4 x \,\mathrm{tr} \Big(
\big\{ \hat{A}_\beta ,
\hat{D}_\mu \hat{D}_\nu \hat{A}_\sigma + 2 \hat{D}_\mu \hat{F}_{\sigma\nu} 
\big\}_\star \star \hat{F}^{\mu\nu} \Big)
\nonumber
\\
& = 
\int d^4 x \,\mathrm{tr} \Big(
\big\{ \hat{A}_\beta ,
-\frac{\mathrm{i}}{2} \big[\hat{F}_{\mu\nu}, \hat{A}_\sigma\big]_\star 
+ \hat{D}_\sigma \hat{F}_{\mu\nu} 
\big\}_\star \star \hat{F}^{\mu\nu} 
\Big)
\nonumber
\\
& = 
\int d^4 x \,\mathrm{tr} \Big(
\frac{\mathrm{i}}{4}  
\big[ \hat{A}_\beta , \hat{A}_\sigma\big]_\star
\star \big\{\hat{F}_{\mu\nu}, \hat{F}^{\mu\nu} \big\}_\star 
- \frac{1}{2} \hat{D}_\sigma \hat{A}_\beta \star 
\big\{\hat{F}_{\mu\nu} , \hat{F}^{\mu\nu} \big\}_\star \Big)~,
\end{align}
where we have used the Bianchi identity and integrated by parts.
Antisymmetrizing in $\beta,\sigma$ we obtain
\begin{align}
&\int d^4 x \,\mathrm{tr} \Big(
\big\{  \hat{A}_\beta ,
\hat{D}_\mu(\partial_\sigma \hat{A}_\nu +  \hat{F}_{\sigma\nu}) 
\big\}_\star \star \hat{F}^{\mu\nu} 
-
\big\{  \hat{A}_\sigma ,
\hat{D}_\mu (\partial_\beta \hat{A}_\nu +  \hat{F}_{\beta\nu}) 
\big\}_\star \star \hat{F}^{\mu\nu} 
\Big)
\nonumber
\\*
& = \int d^4 x \,\mathrm{tr} \Big( 
- \frac{1}{2} \big\{ \hat{F}_{\sigma\beta} , \hat{F}_{\mu\nu} \big\}_\star 
\star \hat{F}^{\mu\nu} \Big)~.
\label{appb4}
\end{align}
Combining (\ref{appb2}) and (\ref{appb4}) we arrive at
\begin{align}
0 &= \int d^4 x \,\mathrm{tr} \Big( 
\hat{D}_\mu \Big(
\frac{1}{4} \big\{ \hat{A}_\beta ,
\partial_\sigma \hat{A}_\nu +  \hat{F}_{\sigma\nu} 
\big\}_\star  
- \frac{1}{4}  \big\{ \hat{A}_\sigma,
\partial_\beta \hat{A}_\nu + \hat{F}_{\beta\nu} \big\}_\star 
+ 2 \frac{d \hat{A}_\nu}{
d \theta^{\beta\sigma}} \Big) 
\star \hat{F}^{\mu\nu} \Big)~,
\end{align}
which leads after reinsertion of $\hat{\Omega}_{\rho\sigma\mu}$ to the
Seiberg-Witten differential equation (\ref{SWA}).

\section{The commutator between rotation and total $\theta$-variation}
\label{appc}

We will prove here eq.\ (\ref{nt}) in the case of rotation.  As usual
it is sufficient to evaluate the commutator on $\hat{A}_\mu$ and on
$\theta^{\mu\nu}$. The last one is zero because rotation and
dilatation of $\theta$ commute, see (\ref{cr}).  In fact the
commutator will vanish for a very general class of differential
equations. Let 
\begin{align} 
\theta^{\rho\sigma}\frac{d \hat{A}_\mu}{d \theta^{\rho\sigma}}
= \theta^{\rho\sigma} \Phi_{\rho\sigma\mu}~,
\end{align}
where $\Phi_{\rho\sigma\mu}$ is a polynomial in\footnote{$\Phi$
  may also depend on the coordinates. In this case however, (\ref{as})
  should also involve rotation of the coordinates.}  $\hat{A}$ and
$\theta$ with power counting dimension 3. We assume that
$\Phi_{\rho\sigma\mu}$ transforms as a tensor under rotation
\begin{align}
W^R_{\hat{A}+\theta;\alpha\beta} \Phi_{\rho\sigma\mu} 
&= \frac{1}{2}\{ x_\alpha,\partial_\beta
\Phi_{\rho\sigma\mu}\}_\star 
-\frac{1}{2}\{ x_\beta,\partial_\alpha \Phi_{\rho\sigma\mu}\}_\star 
\nonumber
\\*
& 
+ g_{\rho\alpha}\Phi_{\beta\sigma\mu} 
- g_{\rho\beta}\Phi_{\alpha\sigma\mu}
+ g_{\sigma\alpha} \Phi_{\rho\beta\mu} 
- g_{\sigma\beta} \Phi_{\rho\alpha\mu} 
+ g_{\mu\alpha}\Phi_{\rho\sigma\beta} 
- g_{\mu\beta}\Phi_{\rho\sigma\alpha}~.
\label{as}
\end{align}
We find
\begin{align}
\Big[W^R_{\hat{A}+\theta;\alpha\beta},\theta^{\rho\sigma} 
\frac{d}{d \theta^{\rho\sigma}}\Big] \hat{A}_\mu 
&= W^R_{\hat{A}+\theta;\alpha\beta}\left(\theta^{\rho\sigma}
\Phi_{\rho\sigma\mu} \right) 
\nonumber
\\
&- \theta^{\rho\sigma} \frac{d}{d\theta^{\rho\sigma}} 
\big( \tfrac{1}{2} \{x_\alpha,\partial_\beta \hat{A}_\mu \}_\star
-\tfrac{1}{2} \{x_\beta,\partial_\alpha \hat{A}_\mu \}_\star
+ g_{\alpha\mu} \hat{A}_\beta - g_{\beta\mu} \hat{A}_\alpha \big)
\nonumber
\\
&=
\theta_\alpha^{~\rho}\left(\Phi_{\rho\beta\mu}
- \Phi_{\beta\rho\mu} \right)
-\theta_\beta^{~\rho} \left( \Phi_{\rho\alpha\mu} 
-\Phi_{\alpha\rho\mu} \right)
\nonumber
\\
&+ \theta^{\rho\sigma} \big( \tfrac{1}{2}\{x_\alpha,
\partial_\beta \Phi_{\rho\sigma\mu} \}_\star 
-                \tfrac{1}{2}\{x_\beta,\partial_\alpha
\Phi_{\rho\sigma\mu} \}_\star 
+ g_{\rho\alpha} \Phi_{\beta\sigma\mu}
- g_{\rho\beta} \Phi_{\alpha\sigma\mu}
\nonumber
\\
& + g_{\sigma\alpha} \Phi_{\rho\beta\mu} - g_{\sigma\beta} 
\Phi_{\rho\alpha\mu}
+ g_{\mu\alpha} \Phi_{\rho\sigma\beta} 
-g_{\mu\beta}\Phi_{\rho\sigma\alpha} \big)
\nonumber
\\
&
-\theta^{\rho\sigma} \left( 
\tfrac{1}{2} \{ x_\alpha, \partial_\beta \Phi_{\rho\sigma\mu} \}_\star 
- \tfrac{1}{2} \{ x_\beta, \partial_\alpha \Phi_{\rho\sigma\mu}
\}_\star 
+g_{\alpha\mu} \Phi_{\rho\sigma\beta}
- g_{\beta\mu} \Phi_{\rho\sigma\alpha}\right)
\nonumber
\\
&=0~.
\end{align}
Now, one checks that $\frac{d \hat{A}_\mu}{d \theta^{\rho\sigma}}$
from (\ref{SWA}) fulfills (\ref{as}), whereby we have proven
(\ref{nt}) for rotation. The proof of (\ref{nt}) in the case of
dilatation is performed in a similar manner. The translational proof
is immediate.

We stress, however, that (\ref{nt}) by no means singles out the
Seiberg-Witten differential equation.

\end{appendix}

\addcontentsline{toc}{section}{\numberline{}References}

\end{document}